\documentclass[aps,pra,showpacs,endfloats]{revtex4}
\usepackage{amsfonts,amsmath,graphicx}
\newcommand{\nc}{\newcommand}
\nc{\rnc}{\renewcommand}
\nc{\bra}[1]{\left\langle {#1} \right|}
\nc{\bbra}[1]{\bra{\mathbf{#1}}}
\nc{\ket}[1]{\left| {#1} \right\rangle}
\nc{\bket}[1]{\ket{\mathbf{#1}}}
\nc{\proj}[1]{\ket{#1}\bra{#1}}
\nc{\sca}[2]{\left\langle #1 | #2 \right\rangle}
\nc{\ketbra}[2]{|#1\rangle\langle#2}
\nc{\half}{\frac{1}{2}}
\nc{\Tr}{\operatorname{Tr}}
\nc{\s}{\mathcal{S}}
\nc{\id}{\leavevmode\hbox{\small1\kern-3.8pt\normalsize1}}

\newtheorem{property}{Property}

\begin{document}

\title{The Geometry of Single-Qubit Maps}

\author{Daniel Kuan Li Oi}
\email[]{daniel.oi@qubit.org}
\affiliation{Centre for Quantum Computation, Clarendon Laboratory, University of Oxford, OX1 3PU, UK}

\date{\today}

\begin{abstract}
The physically allowed quantum evolutions on a single qubit can be described in terms of their geometry. From a
simple parameterisation of unital single-qubit channels, the canonical form of all such channels can be given. The
related geometry can be used to understand how to approximate positive maps by completely-positive maps, such as in the
case of optimal eavesdropping strategies. These quantum channels can be generated by the appropriate network or
through dynamical means. The Str{\o}mer-Woronowisc result can also be understood in terms of this geometry.
\end{abstract}


\maketitle

\section{Introduction}

An important consideration in the field of quantum information theory is what transformations to the state of a
system are physically allowed. This guides us in the search for how we can manipulate quantum information as well
as giving an insight into the nature of quantum theory. It is instructive to examine the simplest non-trivial
case of a quantum system, the qubit, or two-level system, and study the possible evolutions of the state. Such
transformations are called single-qubit channels and they play a vital role in the theory of quantum communication
and computing.  Recently, other workers have analysed the single-qubit channel~\cite{Fujiwara1999} and characterised
the extreme points for the most general qubit map~\cite{Ruskai2001}. The geometry of single-qubit channels provides
a useful description of them. This simple structure of single qubit maps can be applied to other problems in
quantum information theory.

\section{Impossible Qubit Transformations}

If we consider an arbitrary pure state of a qubit, $\ket{\psi}=\alpha\ket{0}+\beta\ket{1}$, what operations are
physically allowed by quantum mechanics? If we require that the final state of the qubit be pure and
non-degenerate, then the operation must be unitary. Therefore, seemingly innocuous operations such as,
\begin{equation} \label{ePositive}
\begin{array}{cc}
\ket{\psi} \mapsto \alpha^* \ket{0} + \beta^* \ket{1}, & \ket{\psi} \mapsto \alpha^* \ket{0} - \beta^* \ket{1}, \\
\ket{\psi} \mapsto \beta^* \ket{0} + \alpha^* \ket{1}, & \ket{\psi} \mapsto \beta^* \ket{0} - \alpha^* \ket{1},
\end{array}
\end{equation}
are not allowed. More generally, if the inital and final states are allowed to be mixed, then the states are
represented by $2$ by $2$ density matrices and the allowed transformations are specific types of maps from density
matrices to density matrices.  These operations are precisely the trace-preserving completely positive (CP) linear
maps.  A map $\s$ is trace-preserving if $\Tr(\s(\rho)) = \Tr(\rho)$ for all density matrices $\rho$ and positive
if the eigenvalues of $\s(\rho)$ are nonnegative  whenever the eigenvalues of $\rho$ are non-negative. This ensures
$\s$ always sends density matrices to density matrices.

The maps in (\ref{ePositive}) are trace-preserving and positive but they fail the final, more subtle,
requirement, complete-positivity. A map $\s$ is said to be completely-positive (CP) if and only if the trivial extension,
$\id_{n}\otimes\s$, is a positive map for all $n$ where $\id_{n}$ is the identity map on $n$ by $n$ matrices. This
requirement is physically very natural as it acknowledges that the mixed state $\rho$ may be entangled with other
quantum systems and that $\s$ extended to these other systems must still produce a physical state. Henceforth, we
will refer to a trace-preserving completely-positive map as a quantum channel.

\section{CP Maps on Single-Qubits}

It will be convenient to work with the Bloch vector representation where the action of a quantum channel on a
single-qubit is characterised by an affine map on the Bloch Sphere:
\begin{equation}
\vec{\text{s}}' = \s(\vec{\text{s}}) = \text{A}\vec{\text{s}}+\vec{\text{b}},
\end{equation}
where $\text{A}$ is a $3\times 3$ real matrix, $\vec{\text{b}}$ is a 3-dimensional real vector and $\vec{\text{s}}$ and $\vec{\text{s}}'$ are
the Bloch vectors representing the initial $\rho=\half(\id+\vec{\text{s}}\cdot\vec{\sigma})$ and final state
$\rho'=\half(\id+\vec{\text{s}}'\cdot\vec{\sigma})$ of the qubit, respectively. Note that such maps are automatically
trace-preserving. We can visualise the effect of these map, $\vec{\text{s}}\mapsto\vec{\text{s}}'$. First, physical states should
be mapped onto physical states and thus $\s$ is a  contraction. Since $\s$ is affine linear, it must map the Bloch
sphere to an ellipsoid contained within the Bloch sphere. This gives 12 free parameters consisting of 6 parameters
denoting the magnitude and axes of scaling of the Bloch sphere, 3 parameters specifying the axis and magnitude of a
rotation and finally, 3 parameters specifying a translation of the ellipsoid. It was shown in \cite{Fujiwara1999}
that not all ellipsoids in the Bloch sphere can be the image of a quantum channel.

If we restrict ourselves to the maps for which $\vec{\text{b}}=0$ (unital), and $\text{A}$ diagonal with entries
$(\eta_{\text{x}},\eta_{\text{y}},\eta_{\text{z}})=\vec{\eta}$, several groups~\cite{Fujiwara1999,Ruskai1999,diVincenzoprivate} have
found neccessary and sufficient conditions for complete positivity. This form is not a serious
restriction as we shall see later. We denote the set of $\vec{\eta}$ corresponding
to CP-maps as $\mathcal{D}$.
\begin{property} $\eta \in \cal{D}$ if and only if
\begin{align}
\label{eCPconds}
\left| \eta_{\text{x}} \pm \eta_{\text{y}} \right| & \leq \left| 1 \pm \eta_{\text{z}} \right|
\end{align}
\end{property}
These conditions specify a tetrahedron in the parameter space of $\{\eta_{\text{x}},\eta_{\text{y}},\eta_{\text{z}}\}$.

\begin{figure}
\centerline{\includegraphics[width=0.5\textwidth]{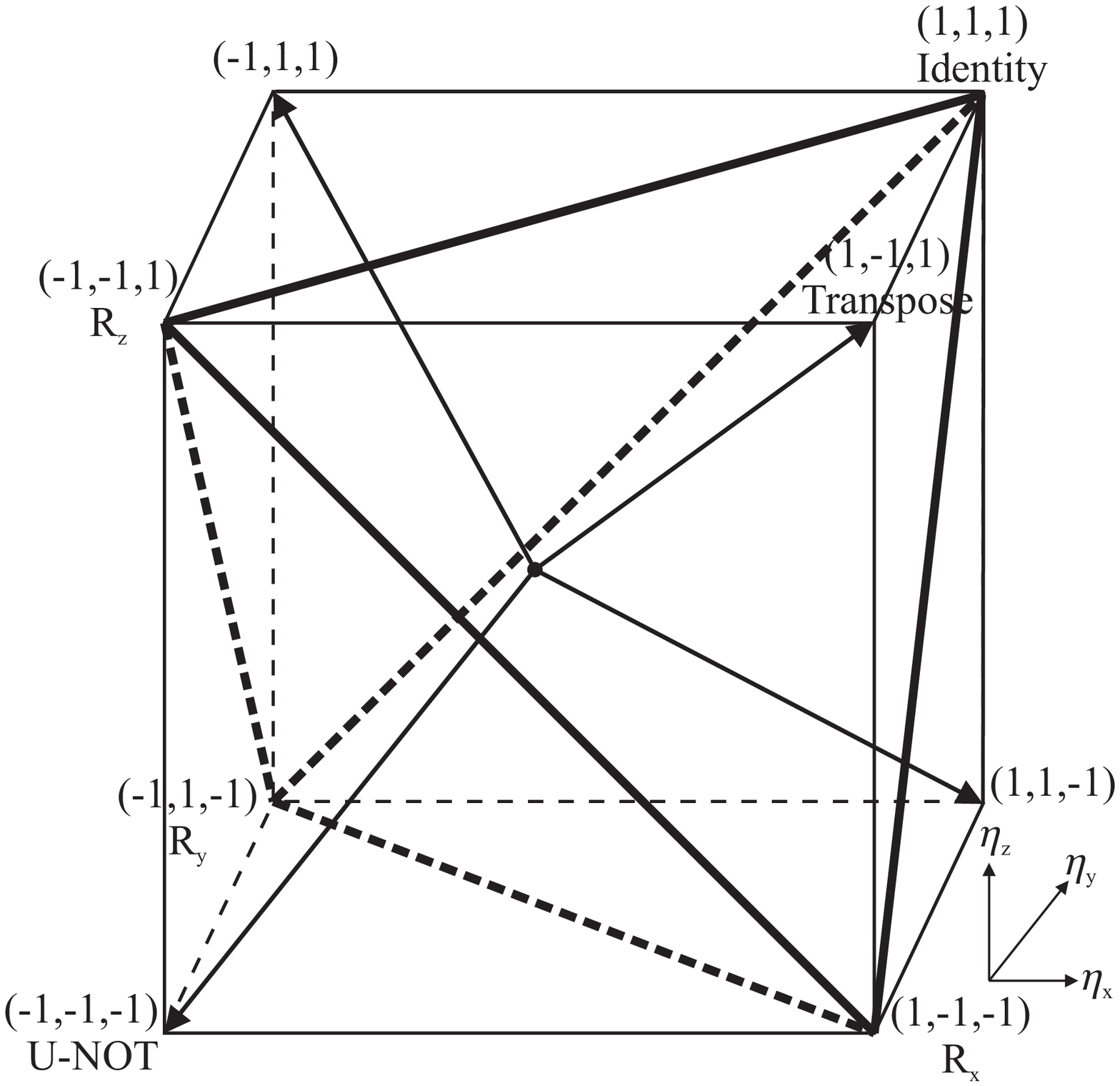}}
\caption[Allowed CP Maps]{The completely positive unital, diagonal single-qubit maps as
a subset of the positive ones.  The axes $\eta_{\text{x}}$, $\eta_{\text{y}}$, and $\eta_{\text{z}}$ represent the diagonal ``squeezing
parameters''.  Some special transformations, including the matrix transpose, the universal NOT, and the identity
are also marked.}
\label{fTetra}
\end{figure}

A simple method of deriving these conditions for unital CP-maps on single-qubits is by examining the effect of a
positive map extended to a maximally entangled system of two qubits, $\ket{\Psi_{+}}=\sum_{i}\ket{i}\ket{i}$. It
can be shown (Appendix) that a necessary and sufficient condition for $\s$ to be a CP-map is that
$\id\otimes\s(\proj{\Psi_{+}})\geq 0$. For a unital diagonal map
$\s$, this leads to,
\begin{equation}
\rho'=\frac{1}{2}\left(
\begin{array}{cccc}
1+\eta_{\text{z}} & 0 & 0 & \eta_{\text{x}}+\eta_{\text{y}} \\
0 & 1-\eta_{\text{z}} & \eta_{\text{x}}-\eta_{\text{y}} & 0 \\
0 & \eta_{\text{x}}-\eta_{\text{y}} & 1-\eta_{\text{z}} & 0 \\
\eta_{\text{x}}+ \eta_{\text{y}} & 0 & 0 & 1+\eta_{\text{z}}
\end{array}\right)
\end{equation}
being positive when,
\begin{subequations}
\begin{eqnarray}
(1+\eta_{\text{z}})^2-(\eta_{\text{x}}+\eta_{\text{y}})^2 &\geq& 0\\
(1-\eta_{\text{z}})^2-(\eta_{\text{x}}-\eta_{\text{y}})^2 &\geq& 0.
\end{eqnarray}
\end{subequations}
These are precisely the tetrahedron conditions for the unital single-qubit channel. The geometry of $\cal{D}$ in
this representation  will be the starting point for our investigations (Figure \ref{fTetra}).

\section{Simulating Quantum Channels} \label{sNetwork}

The set $\cal{D}$ is a regular tetrahedron with vertices $\text{I}$,
$\text{R}_{\text{x}}$, $\text{R}_{\text{y}}$ and $\text{R}_{\text{z}}$, where $\text{I}$ is
the identity tranformation and the $\text{R}_{i}$s are rotations by $\pi$ about the $\text{x}$, $\text{y}$, $\text{z}$ axes. Since the
tetrahedron is a convex polyhedron, $\cal{D}$ is the convex hull of the
points representing $I$ and the three rotations.  Thus, every transformation corresponding to a point in
$\cal{D}$ can be realised as a  statistical mixture of those four extremal transformations.  Such a transformation
is sometimes referred to as a Pauli channel. Thus, we have
\begin{property}
$\vec{\eta} \in \cal{D}$ if and only if $\vec{\eta}$ corresponds
to a Pauli channel.
\end{property}

\begin{figure}
\centerline{\includegraphics[width=0.5\textwidth]{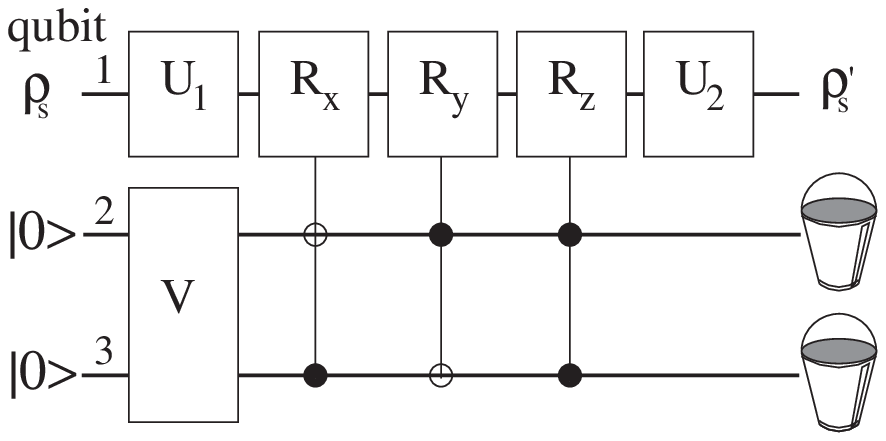}}
\caption[Quantum Network]{A quantum computational network for simulating a unital single-qubit
quantum channel.  Transformations $\text{U}_1$ and $\text{U}_2$ are qubit rotations
while $\text{V}$ prepares arbitrary superpositions of the
states $\ket{00}, \ket{01}, \ket{10}, \ket{11}$ of the ancilla. After interacting with qubit 1, the ancilla is
ignored. The controlled-controlled-$\text{R}_{i}$ gates perform the Pauli rotations depending on the component
amplitudes of the ancilla.}
\label{fNetwork}
\end{figure}

Now, suppose $\s$ is a unital single-qubit quantum channel but $\text{A}$ is not necessarily diagonal. $\text{A}$ possesses a polar
decomposition of the form $\text{A} = \text{s} \text{P} \text{R}$, where $\text{s}=\det(\text{A})$,
$\text{P}=\left(\text{AA}^{\text{T}}\right)^\frac{1}{2}$, and $\text{R}$ is a rotation~\cite{horn}.
Since $\text{sP}$ is symmetric, there exists a rotation $\text{Q}$ and a diagonal $\Delta$ such that
$\text{sP} = \text{Q}\Delta \text{Q}^{\text{T}}$, giving
\begin{equation}
\text{A} = \text{Q} \Delta \text{Q}^{\text{T}} \text{R}.
\end{equation}
Since both $\text{Q}$ and $\text{Q}^{\text{T}} \text{R}$ are rotations, a general unital CP map on a single qubit 
can be decomposed into a rotation, followed by a diagonal transformation, followed by another rotation. Thus, we
can construct a quantum computational network to simulate an arbitrary single-qubit unital quantum 
channel (Figure~\ref{fNetwork}).

\section{CP Map Dynamics}

\begin{figure}
\centerline{\includegraphics[width=0.5\textwidth]{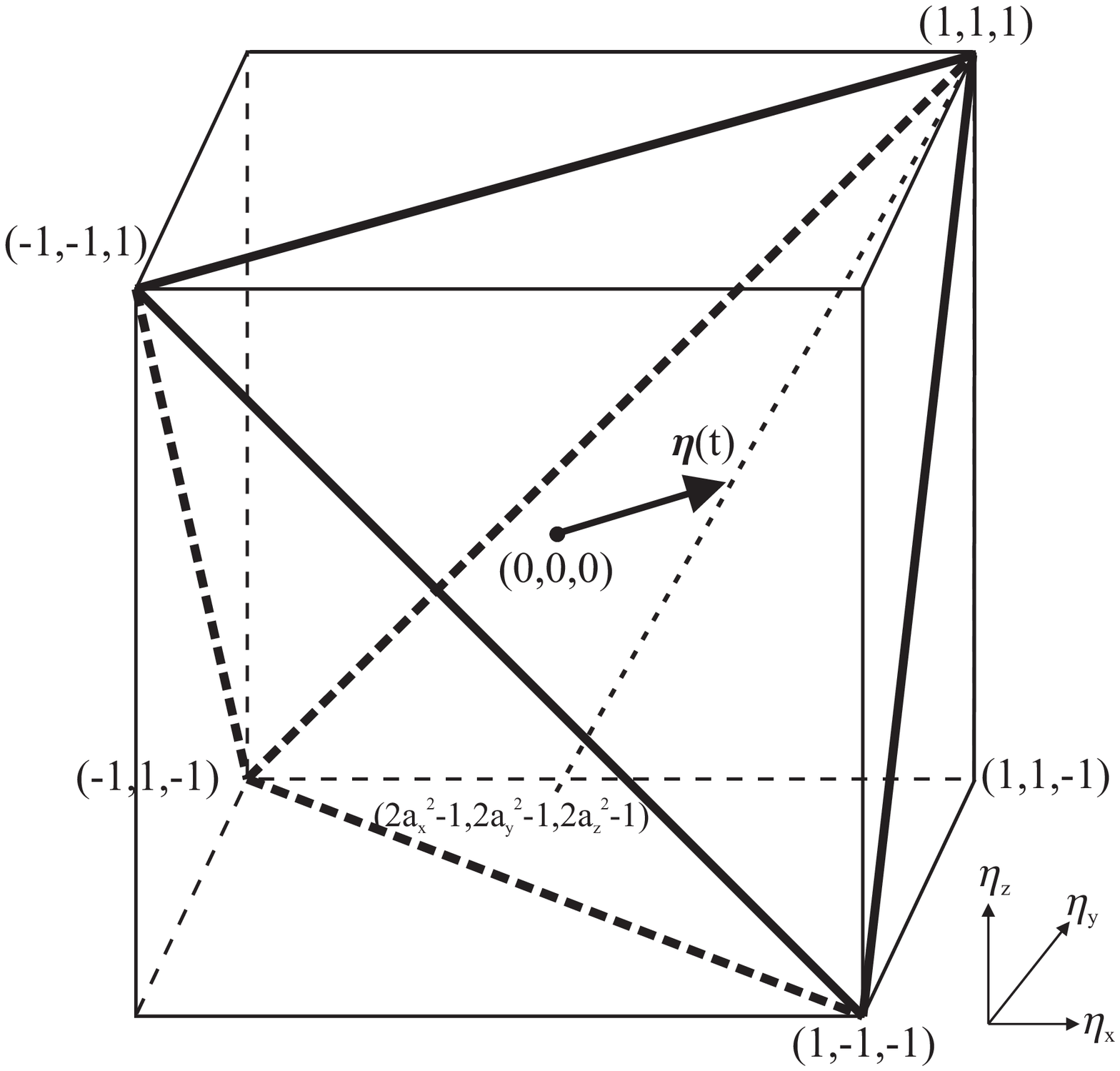}}
\caption[CP Map Dynamics]{Dynamical generation of a CP map by suitable combination of coupling Hamiltonians and evolution times.
\label{fcpmaphamil}}
\end{figure}

\begin{figure}
\centerline{\includegraphics[width=0.95\textwidth]{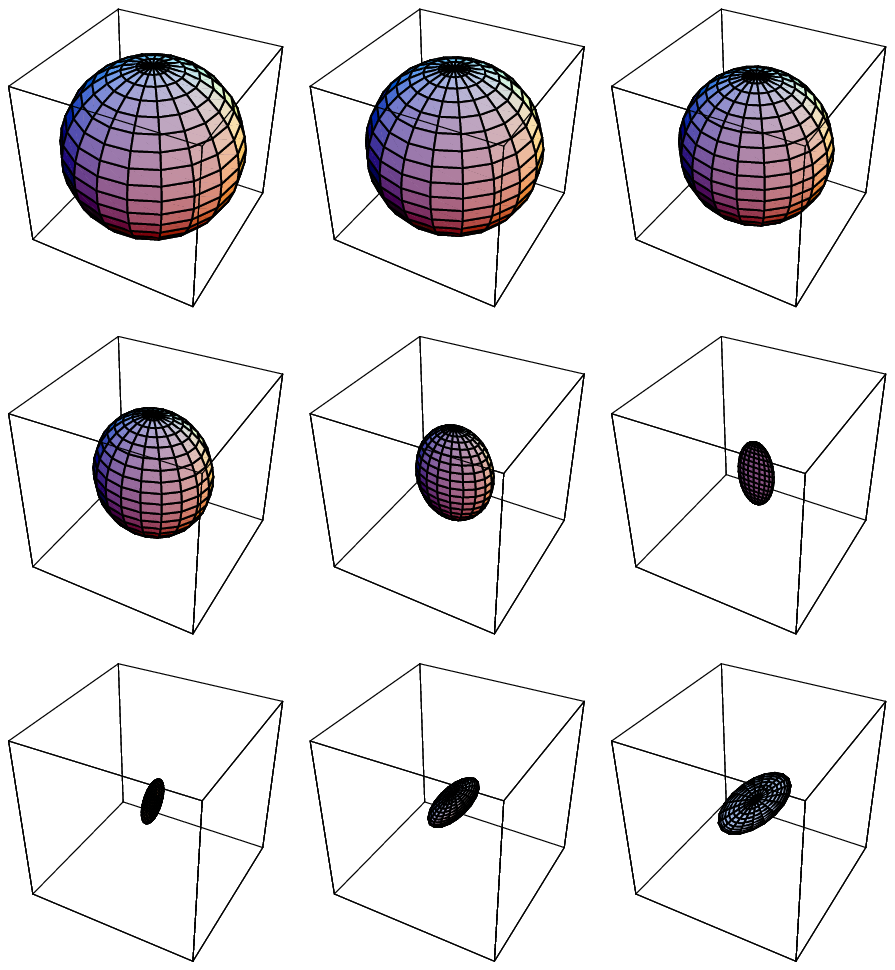}}
\caption[Bloch Sphere Dynamics]{The Bloch sphere of the reduced density operator of the single qubit evolving in time with
$\alpha_{\text{x}}^{2}=\frac{1}{2}$, $\alpha_{\text{y}}^{2}=\frac{1}{3}$ and $\alpha_{\text{z}}^{2}=\frac{1}{6}$.
\label{fbloch}}
\end{figure}

We can simulate a CP map dynamically by coupling the qubit to an ancilla with a time-independant Hamiltonian,
evolving the whole system by the unitary, $\text{U}(\text{t})=\text{exp}\left(\frac{-i\text{Ht}}{\hbar}\right)$, and
tracing over the ancilla. Thus, the final state of the qubit is a function of the interaction
time. This gives us a class of maps on the single-qubit subsystem, parameterised by t, which depends on the coupling
Hamiltonian. For unital maps, these classes corespond to paths in $\vec{\eta}$-space. It is again convenient to use a
two-qubit ancilla. Consider the following Hamiltonians,
\begin{subequations}
\begin{eqnarray}
\text{H}_{\text{x}}&=&\sigma_{\text{x}}\otimes\left(\ket{\text{a}_{1}}\bra{\text{a}_{2}}+\ket{\text{a}_{2}}\bra{\text{a}_{1}}\right)\\
\text{H}_{\text{y}}&=&\sigma_{\text{y}}\otimes\left(\ket{\text{a}_{1}}\bra{\text{a}_{3}}+\ket{\text{a}_{3}}\bra{\text{a}_{1}}\right)\\
\text{H}_{\text{z}}&=&\sigma_{\text{z}}\otimes\left(\ket{\text{a}_{1}}\bra{\text{a}_{4}}+\ket{\text{a}_{4}}\bra{\text{a}_{1}}\right),
\end{eqnarray}
\end{subequations}
where $\left\{\ket{\text{a}_{i}}\right\}$ are othornormal states of the ancilla, and the initial state of the
system is $\rho_{0}\otimes\proj{\text{a}_{1}}$. Each $\text{H}_{i}$ induces a set of CP maps in
$\vec{\eta}$-space which form a straight line from $\text{I}$ to the other corners of the tetrahedron,
$\text{R}_{\text{x}}$, $\text{R}_{\text{y}}$ or $\text{R}_{\text{z}}$ respectively. If we combine these Hamiltonians together,
\begin{eqnarray}
\text{H}_{\text{total}}&=&
\alpha_{\text{x}}\text{H}_{\text{x}}+\alpha_{\text{y}}\text{H}_{\text{y}}+\alpha_{\text{z}}\text{H}_{\text{z}}\nonumber\\
1&=&\alpha_{\text{x}}^{2}+\alpha_{\text{y}}^{2}+\alpha_{\text{z}}^{2},
\end{eqnarray}
the map induced by $\text{H}_{\text{total}}$ is,
\begin{equation}
\vec{\eta}(\text{t})=%
(1,1,1)\cos^2\left(\frac{\text{t}}{\hbar}\right)+%
\left(2\alpha_{\text{x}}^2-1,2\alpha_{\text{y}}^2-1,2\alpha_{\text{z}}^2-1\right)\sin^2\left(\frac{\text{t}}{\hbar}\right)
\end{equation}
This resulting set of maps is a line in $\vec{\eta}$-space connecting $\text{I}$ to the convex combination of
$\text{R}_{\text{x}}$, $\text{R}_{\text{y}}$ and $\text{R}_{\text{z}}$ weighted by the $\alpha_{i}^2$. For example, if
$\alpha_{x}^2=\alpha_{y}^2=\alpha_{z}^2=\frac{1}{3}$, the resulting set of maps is a line from $(1,1,1)$ to
$\left(-\frac{1}{3},-\frac{1}{3},-\frac{1}{3}\right)$. Thus, if we let the system evolve until $\text{t}=\frac{\pi\hbar}{2}$,
the induced map is the best approximation to the Universal-NOT, as we shall see later. At
$\text{t}=\frac{\pi\hbar}{3},\frac{2\pi\hbar}{3}$, the qubit is maximally mixed. Conversely, any unital CP map can
be expressed as the result of some combination of $\text{H}_{\text{x}}$, $\text{H}_{\text{y}}$ and
$\text{H}_{\text{z}}$ evolved for a particular time (Figures~\ref{fcpmaphamil}~and~\ref{fbloch}).

\section{Approximating Positive Maps}

\begin{figure}
\centerline{\includegraphics[width=0.5\textwidth]{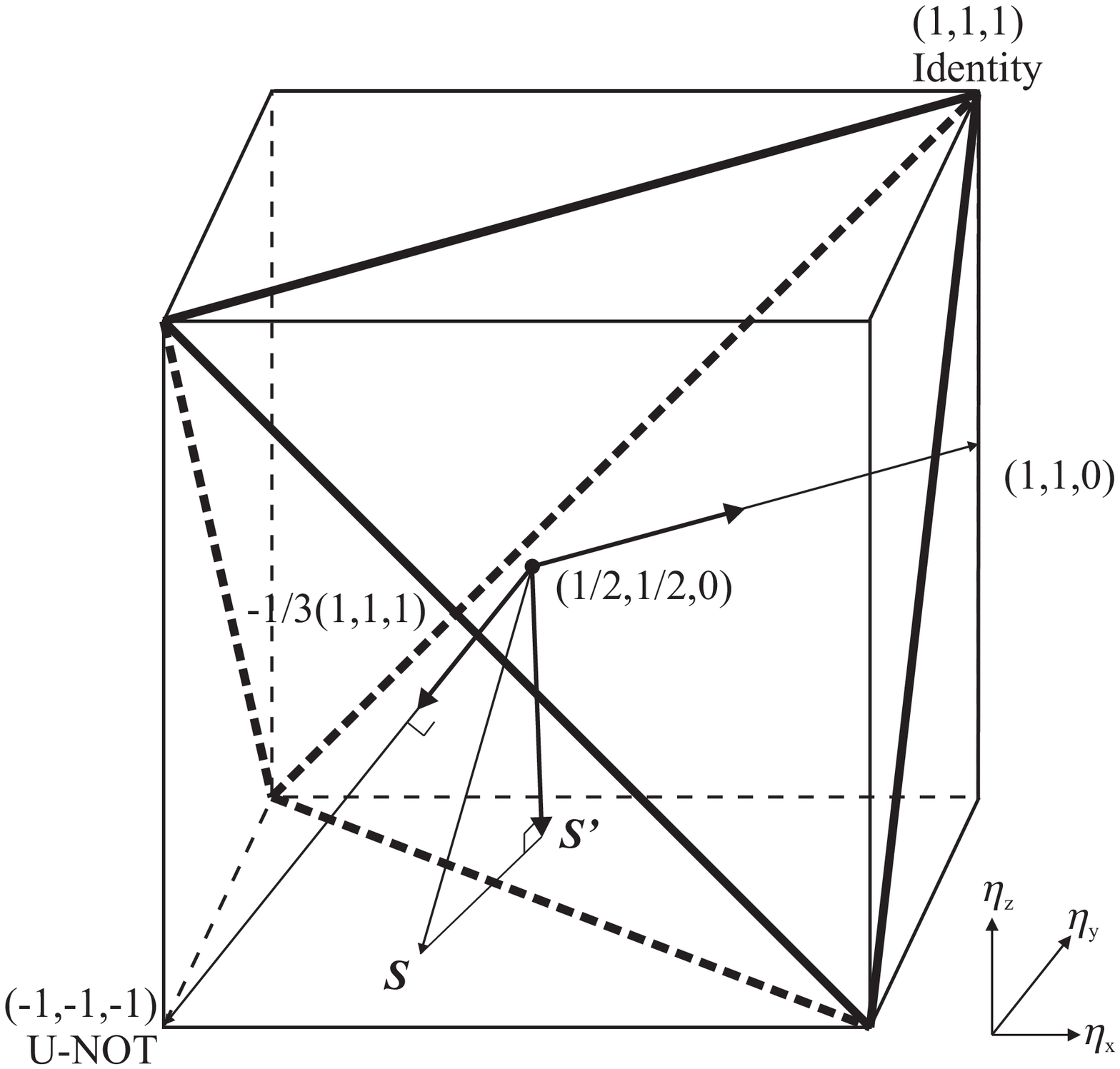}}
\caption[Approximating Maps]{Both the universal NOT and a generic positive map are approximated
by dropping perpendiculars to the closest face of $\cal{D}$.}
\label{fApprox}
\end{figure}

Now consider finding the CP map that best approximates a given positive map on
a single qubit. We shall choose the metric on the space of positive maps induced by the inner product,
\begin{equation}
\langle \text{A},\text{B} \rangle = \text{Tr}(\text{A}^{\dagger}\text{B})
\end{equation}
For the unital diagonal maps, this is simply the Euclidean distance in $\vec{\eta}$-space. Thus, finding
the best unital, diagonal CP approximation to a given transformation is simply a matter of minimising the distance
to the tetrahedron $\mathcal{D}$, in effect, dropping a perpendicular to the nearest face of $\mathcal{D}$
(Figure~\ref{fApprox}). This simplification has wide applicability because symmetry requirements on the
approximating maps often necessitate the restriction to unital, diagonal maps anyway.

A simple example is the best approximation to the universal NOT gate on a single qubit~\cite{buzek}.  On
the Bloch sphere, a perfect universal NOT corresponds to the transformation $\vec{\eta}=(-1,-1,-1)$.  It is simply the map
\begin{equation}
\ket{\psi} \mapsto \beta^{*}\ket{0} - \alpha^{*}\ket{1}
\end{equation}
that we observed was not physical in (\ref{ePositive}). Here, we need to drop a perpendicular to the plane
\begin{equation}
\eta_{\text{x}} + \eta_{\text{y}} + \eta_{\text{z}} = -1.
\end{equation}
This yields the best approximation, $\vec{\eta}=(-\frac{1}{3},-\frac{1}{3},-\frac{1}{3})$ which was, of course,
well-known.  It is easy to check that this map can be constructed by selecting randomly from among the three
options $\text{R}_{\text{x}}$, $\text{R}_{\text{y}}$, and $\text{R}_{\text{z}}$, each with probability one third.

The technique, however, applies equally well to less symmetrical positive maps not as amenable to the techniques
used in \cite{buzek}. For example, the best approximation to the ``pancake'' map  $\vec{\eta}=(1,1,0)$ is given by
the map $\vec{\eta}=(\frac{2}{3},\frac{2}{3},\frac{1}{3})$.  It is also easy to verify that the best
approximation of the type $(\eta_{\text{x}},\eta_{\text{y}},0)$ is  $\vec{\eta}=(\frac{1}{2},\frac{1}{2},0)$.

\section{Application to Quantum Eavesdropping}

In the \emph{four-state}~\cite{BB84} (resp. \emph{six-state}~\cite{Bruss}) quantum key-distribution protocol, Alice
sends one by one to Bob, qubits in one of four (six) states, $\{\ket{0}_\theta,\ket{1}_\theta\}_\theta$ where
$\theta\in\{\text{x},\text{z}\}$ ($\theta\in\{\text{x},\text{y},\text{z}\}$) is a choice of basis and 
$\ket{0}_{\text{z}}=\ket{0}$, $\ket{1}_{\text{z}}=\ket{1}$,
$\ket{0}_{\text{x}}=(\ket{0}+\ket{1})/2$, $\ket{1}_{\text{x}}=(\ket{0}-\ket{1})/2$, $\ket{0}_{\text{y}}=(\ket{0}+ i\ket{1})/2$ and
$\ket{1}_{\text{y}}=(\ket{0}-i \ket{1})/2$. Bob measures each qubit in a random basis $\theta'\in\{\text{x},\text{z}\}$
($\theta'\in\{\text{x},\text{y},\text{z}\}$) chosen independently of $\theta$. After the quantum transmission, Alice and Bob compare
their bases publicly. For each qubit, if their bases correspond, Alice and Bob should share the same bit value,
provided the qubits were not tampered with during the transmission. They estimate the error rate or the disturbance
of the quantum channel by comparing a sample of these bits. The key distribution is validated if this disturbance
is smaller than a certain specified threshold. 

Many eavesdropping scenerios against quantum cryptographic protocols have been studied. In
particular, in an \emph{incoherent} attack, a possible spy, Eve, interacts each qubit, sent by Alice to Bob, with a
probe. One can assume, without loss of generality, that the probe is in a pure state $\ket{\text{E}}$. The unitary
operator that Eve uses to interact her probe with Alice's qubits is identical in each case. In the basis $\theta\in\{\text{x},\text{z}\}$
(or $\theta\in\{\text{x},\text{y},\text{z}\}$)
\begin{eqnarray}\label{eavesdrop1}
\text{U}\ket{0}_\theta\ket{\text{E}} &=& \ket{\text{E}_{0 0}}_\theta \ket{0}_\theta + \ket{\text{E}_{0 1}}_\theta \ket{1}_\theta\\
\text{U}\ket{1}_\theta\ket{\text{E}} &=& \ket{\text{E}_{1 0}}_\theta \ket{0}_\theta + \ket{\text{E}_{1 1}}_\theta
\ket{1}_\theta
\label{eavesdrop2},
\end{eqnarray}
where the $\ket{\text{E}_{i j}}_\theta$ are possibly not normalised nor orthogonal states for Eve's probe. After
transmission of qubits, Eve stores her probes until the public announcement of the bases and measures them
accordingly. However, Eve is limited to measuring her probes individually. In a \emph{symmetric incoherent} attack,
the quantum channel described above is a Pauli channel $(\eta_{\text{x}},\eta_{\text{y}},\eta_{\text{z}})$ with $\eta_{\text{x}}=\eta_{\text{z}}=\eta$
for the four-state protocol and $\eta_{\text{x}}=\eta_{\text{y}}=\eta_{\text{z}}=\eta$ for the six-state protocol. This implies that, for any
basis $\theta$,
\begin{eqnarray}
\,_\theta\langle \text{E}_{0 1} | \text{E}_{0 1} \rangle_\theta &=&
\,_\theta\langle \text{E}_{1 0} | \text{E}_{1 0} \rangle_\theta = \frac{1-\eta}{2}=\text{D}\\
\,_\theta\langle \text{E}_{0 0} | \text{E}_{0 0} \rangle_\theta &=&
\,_\theta\langle \text{E}_{1 1} | \text{E}_{1 1} \rangle_\theta = \frac{1+\eta}{2}=\text{F}=1-\text{D}\\
\,_\theta\langle \text{E}_{0 0} | \text{E}_{1 1} \rangle_\theta &=&
\,_\theta\langle \text{E}_{1 1} | \text{E}_{0 0} \rangle_\theta = \left\{
  \begin{array}{cl}
   (\eta+\eta_{\text{y}})/2 \quad & \text{4-state}\\ 
    \eta \quad & \text{6-state}
  \end{array}\right.
\end{eqnarray}

The quantity $\text{D}$, called \emph{disturbance}, is the probability, given that they choose the same basis, Alice
and Bob get a different bit. Similarly, the quantity $\text{F}$ is called \emph{fidelity} since it is the probability,
given Alice and Bob choose the same basis, they get the same bit. Let $\text{p}_{\text{c}}$ be the probability that given
Alice and Bob share the same basis and the same bit, Eve guesses correctly the value of this shared bit. When Alice
and Bob share the same basis and the same bit, Eve has to guess whether her probe is in state
$\frac{1}{\text{F}}\ket{\text{E}_{0 0}}_{\theta\, \theta}\bra{\text{E}_{0 0}}$ or in the state
$\frac{1}{\text{F}}\ket{\text{E}_{1 1}}_{\theta\,\theta}\bra{\text{E}_{1 1}}$.
If Eve uses the optimal measurement to guess this bit value,
\begin{equation}
\text{p}_{\text{c}} =
\frac{1}{2}+\frac{1}{2}\sqrt{ 1-\frac{1}{\text{F}} |\,_\theta\langle \text{E}_{0 0} | \text{E}_{1 1} \rangle_\theta|^2}.
\end{equation}

\begin{figure}
\centerline{\includegraphics[width=0.5\textwidth]{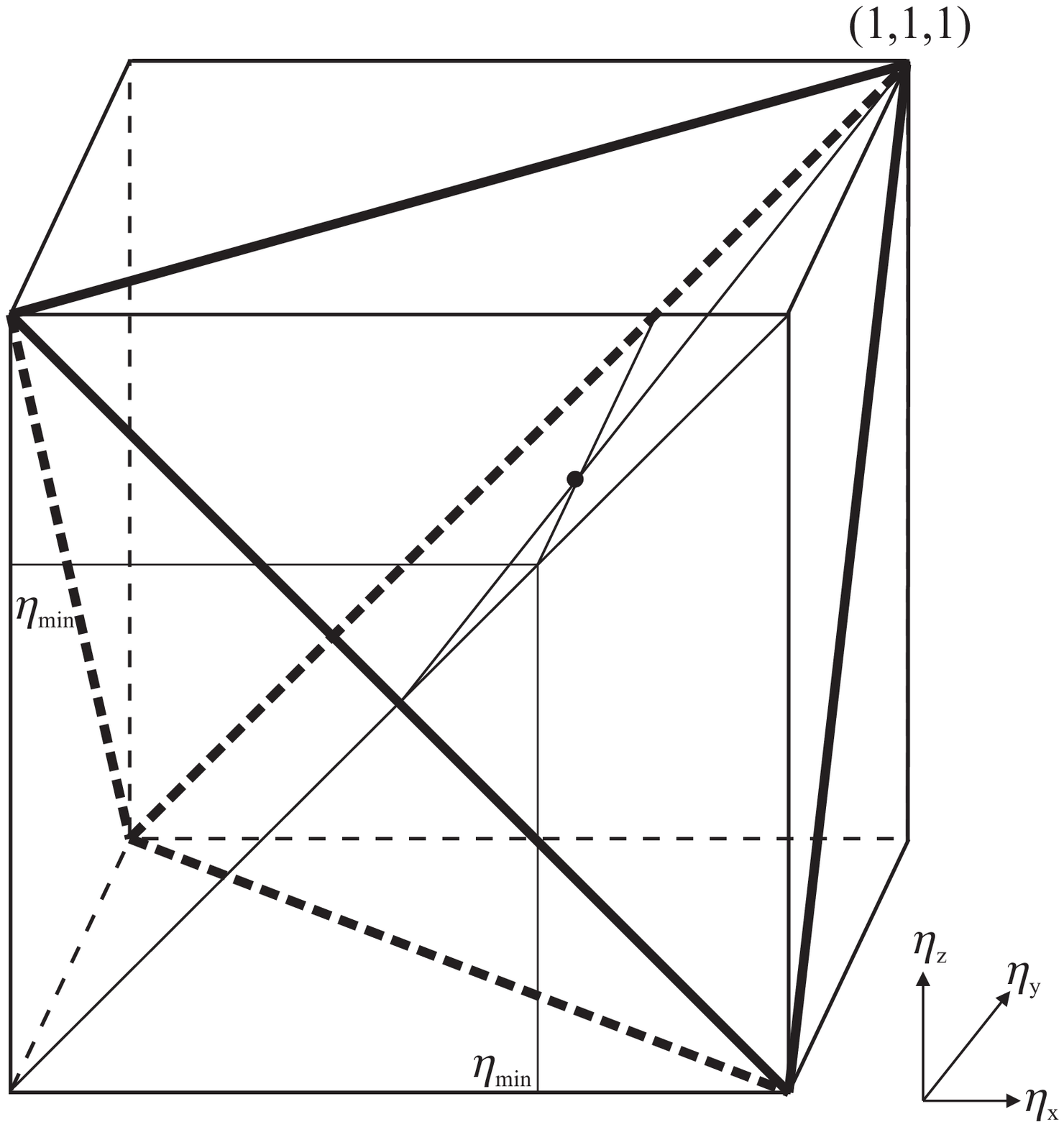}}
\caption[4 State Protocol]{For the four-state protocol with $\eta_{\text{x}}=\eta_{\text{z}}=\eta \geq
\eta_{\text{min}}$, Eve has to minimise $\eta_{\text{y}}$. When $\eta_{\text{min}}$ is positive, the minimum lies on
the plane $\{(1,1,1),(-1,-1,1),(1,-1,1)\}$.}
\label{f4State}
\end{figure}

\begin{figure}
\centerline{\includegraphics[width=0.5\textwidth]{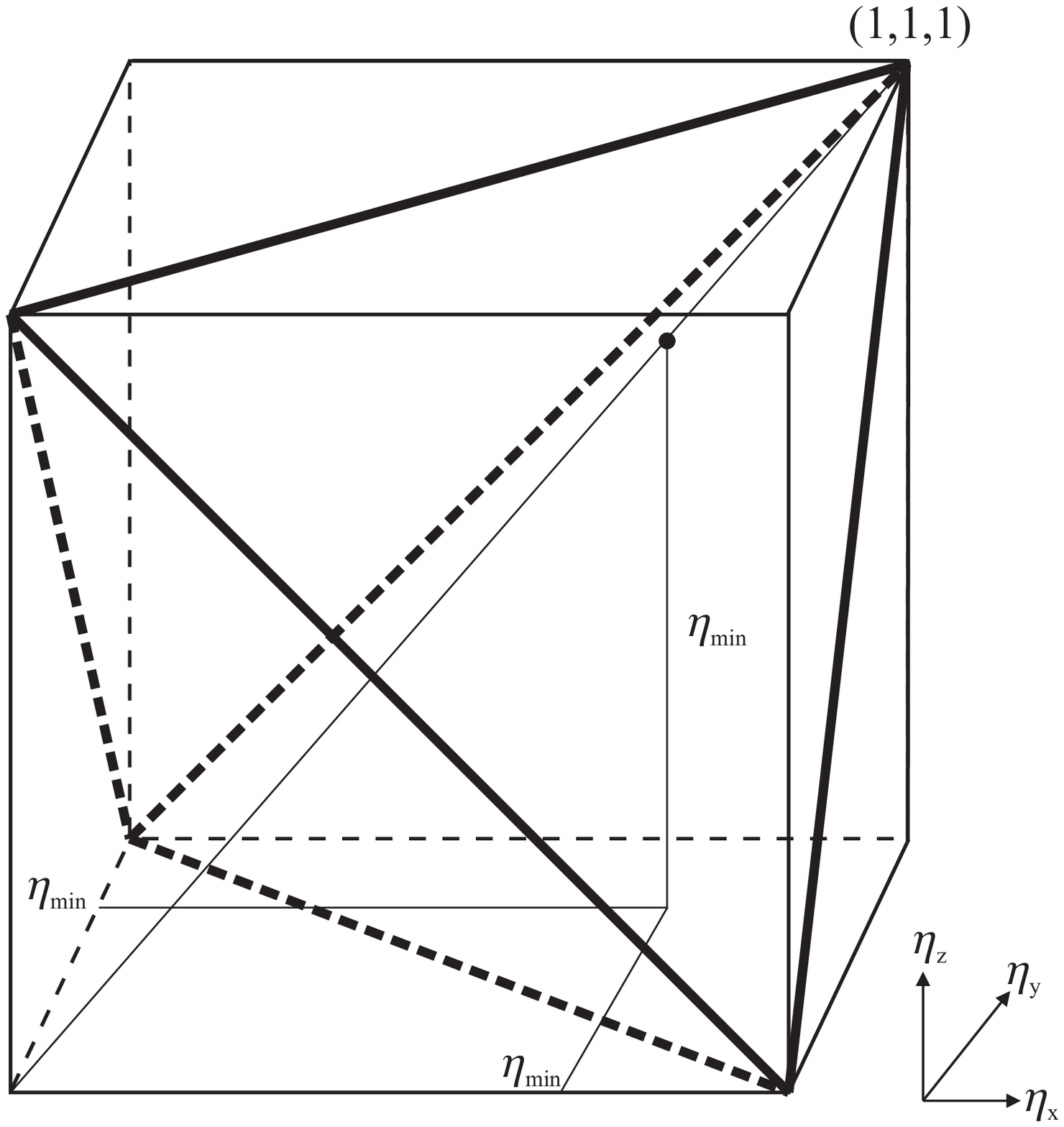}}
\caption[6 State Protocol]{For the six-state protocol, with $\eta_{\text{x}}=\eta_{\text{y}}=\eta_{\text{z}}=\eta \geq \eta_{\text{min}}$,
the minimum is reached for $\eta=\eta_{\text{min}}$.}
\label{f6State}
\end{figure}

Eve's objective is to maximise $\text{p}_{\text{c}}$ given an allowed disturbance level. In other words, given $\eta \geq
\eta_{min}$ where $(1-\eta_{\text{min}})/2=\text{D}_{\text{max}}$ is the maximum allowed disturbance, Eve has to minimise
$|\,_\theta\langle \text{E}_{0 0} | \text{E}_{1 1} \rangle_\theta|$. Referring to the tetrahedron representing $\mathcal{D}$, it
is easy to see that such a minimum is reached for $\vec\eta = (\eta_{\text{min}},2 \eta_{\text{min}}-1, \eta_{\text{min}})$ for the
four-state protocol (Figure~\ref{f4State}) and $\vec\eta = (\eta_{\text{min}},\eta_{\text{min}}, \eta_{\text{min}})$ for the six-state
protocol (Figure~\ref{f6State}). This is precisely the results obtained by Cirac and Gisin in~\cite{CG}.

It should be noted that the same analysis can be applied to protocols involving distribution of EPR pairs between
Alice and Bob~\cite{Ekert2000} and, in essence, the same results apply.

\section{Str{\o}mer-Woronowicz Result} \label{sTranspose}

Horodecki's proof \cite{horodecki} of Peres' separability criterion \cite{peres} for 2 by 2 and 2 by 3
dimensional quantum systems relies crucially on an older result of Str{\o}mer and Woronowicz \cite{stormer} stating
that any positive map $\mathcal{P}$ from a 2 dimensional quantum system to a 2 or 3 dimensional quantum system can
be decomposed in the form
\begin{equation}
\label{eStormer}
\mathcal{P} = \mathcal{CP}_{1} + \mathcal{CP}_{2} \circ \mathcal{T},
\end{equation}
where $\mathcal{CP}_{1}$ and $\mathcal{CP}_{2}$ are completely positive maps and $\mathcal{T}$ is the transpose
map.  The geometry of $\cal{D}$ will again make the reason clear for the 2 by 2 dimensional, unital case.

The maps in (\ref{eStormer}) are not neccessarily trace preserving so it is convenient to consider an equivalent
result,
\begin{equation}
\mathcal{P} = \text{p}\mathcal{CP}_{1} + (1-\text{p})\mathcal{CP}_{2} \circ \mathcal{T},
\end{equation}
where $0\leq \text{p}\leq 1$ and $\mathcal{CP}_1$ and $\mathcal{CP}_2$ are now trace preserving. Thus,
$\mathcal{P}$ can be expressed as the convex combination of $\mathcal{CP}_1$ and $\mathcal{CP}_2\circ\mathcal{T}$.

Furthermore, observe that $\mathcal{T}$ corresponds to the point $(1,-1,1)$ in Figure \ref{fTetra} and is
equivalent to the points $(-1,-1,-1)$, $(-1,1,1)$ and $(1,1,-1)$ up to rotations of the tetrahedron. Since
rotations correspond to unitary transformations, which are invertible CP maps, any of the above 3 transformations
can be substituted for $\mathcal{T}$ in (\ref{eStormer}).

Now, given any point inside the cube of Figure~\ref{fTetra} representing a positive map $\mathcal{P}$, it either
lies within $\mathcal{D}$, or within one of the four pyramidal regions. If it lies within the tetrathedron, then
the proof is trivial. If the map lies outside $\mathcal{D}$, the map is positive but not completely positive and we
need to show that it can be decomposed into $\mathcal{CP}_{1}$ and $\mathcal{CP}_{2}\circ \mathcal{T}$, where
$\mathcal{CP}_{1}$ and $\mathcal{CP}_{2}$ both lie within $\mathcal{D}$.

\begin{figure}
\centerline{\includegraphics[width=0.5\textwidth]{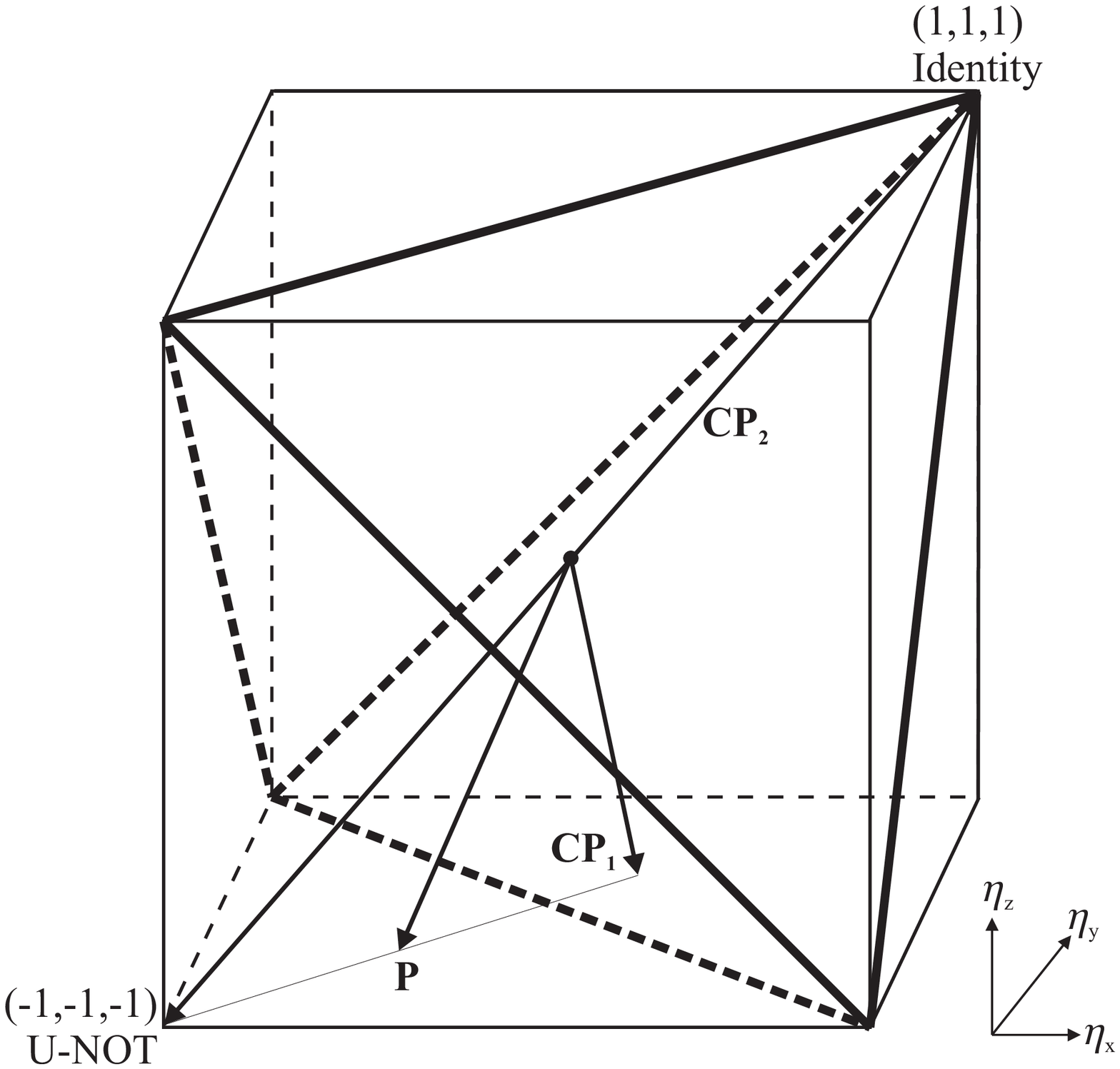}}
\caption[Stromer-Woronowisc]{The map $\mathcal{P}$ is decomposed into components $\mathcal{CP}_{1}$ and
$\mathcal{CP}_{2}\circ\text{U-NOT}$ lying entirely within $\cal{D}$.}
\label{fStormer}
\end{figure}

To simplify the argument, let us consider the corner region of the U-NOT map $(-1,-1,-1)$. By symmetry, the following
argument can be applied to the other three corners regions. By constructing a line between the map $(-1,-1,-1)$ and
the face of the tetrahedron whose vertices are $\text{R}_{\text{x}}$, $\text{R}_{\text{y}}$ and $\text{R}_{\text{z}}$,
passing through the map $\mathcal{P}$, the proof is now apparent. Every $\mathcal{P}$ can be decomposed in this way
as the non-CP pyramidal region is convex with extreme points, $(-1,-1,-1)$, $\text{R}_{\text{x}}$,
$\text{R}_{\text{y}}$ and $\text{R}_{\text{z}}$. Figure \ref{fStormer} illustrates the situation.

\section{Conclusion}

We have seen how the tetrahedral geometry of the set of unital, diagonal
single-qubit channels can be used to motivate solutions to a number
of problems in quantum information theory.  The question of whether the
techniques can be extended to non-unital maps or to higher dimensions
remains open.

\begin{acknowledgments}
I would like to thank Artur Ekert, Patrick Hayden and Hitoshi Inamori for illuminating discussions and valuable
feedback on this work. I would also like to acknowledge the support of CESG, UK.
\end{acknowledgments}

\appendix

\section*{Appendix}

The CP conditions~(\ref{eCPconds}) can be easily derived by ensuring that the trivial extension of
a positive map to a maximally entangled state is positive. Let $\s :
\mathcal{B}(C^d) \mapsto \mathcal{B}(C^d)$ be a linear operator and let
$\ket{\Psi_{+}}=\text{N}^{-\frac{1}{2}}\sum_{i}\ket{i}\ket{i}$ be a maximally entangled state. Then
\begin{equation}
(\id\otimes\s)\proj{\Psi_{+}} \geq 0,
\end{equation}
iff $\s$ is CP. To see this, if $\s$ is not CP, $\exists\ket{\Phi}\in
\mathcal{H}\otimes\mathcal{H}$ such that $(\id\otimes\s)\proj{\Phi}\not\geq 0$. We can express,
\begin{equation}
\ket{\Phi}=(\mathcal{A}\otimes\id)\ket{\Psi_{+}},
\end{equation}
where $\mathcal{A}$ is CP and its components are,
\begin{eqnarray}
\bra{m}\text{A}\ket{n}&=&\sqrt{\text{N}}a_{mn}\nonumber\\
(\text{A}\otimes\id)\ket{\Psi_{+}}&=&\sum_{m,n}a_{mn}\ket{m}\ket{n}\nonumber\\
\rho&\mapsto&\text{A}\rho\text{A}^{\dagger}.
\end{eqnarray}
Thus, we have that,
\begin{eqnarray}
(\id\otimes\s)(\mathcal{A}\otimes\id)\proj{\Psi_{+}}&\not\geq& 0\nonumber\\
(\mathcal{A}\otimes\id)(\id\otimes\s)\proj{\Psi_{+}}&\not\geq& 0\nonumber\\
\Rightarrow \quad (\id\otimes\s)\proj{\Psi_{+}}&\not\geq& 0,
\end{eqnarray}
hence we only need look at the action of $\id\otimes\s$ on the maximally entangled state $\ket{\Psi_{+}}$
to ensure that $\s$ is CP.

\bibliography{prageom02}

\end{document}